\documentclass[aps,pre,twocolumn,groupedaddress,floatfix]{revtex4}
\usepackage{graphicx}
\usepackage{amsmath}
\begin{document}
\def\deg{^\circ}
\newcommand{\ind}[1]{_{\mbox{\scriptsize#1}}}     
\newcommand{\indt}[1]{_{\mbox{\tiny#1}}}          
\newcommand{\supind}[1]{^{\mbox{\scriptsize#1}}}  
\newcommand{\unit}[1]{\, \mbox{#1}}               
\newcommand{\imag}{\mbox{i}}                      
\newcommand{\imind}{\mbox{\scriptsize{i}}}        

\title{Flow profiling of a surface acoustic wave nanopump}

\author{Z. Guttenberg$^{1}$, A. Rathgeber$^{1}$, S. Keller$^{2}$, J.O.
R\"adler$^{2}$,\\ A. Wixforth$^{3}$, M. Kostur$^3$, M. Schindler$^3$,
P. Talkner$^3$}
\affiliation{$^{1}$Advalytix AG, Eugen-S\"anger-Ring 53.0,  
D-85649 Brunnthal, Germany. \\
$^{2}$Ludwig-Maximilians-Universit\"at, Geschwister-Scholl-Platz
1, D-80539 M\"unchen, Germany.\\
$^3$Institut f\"ur Physik, Universit\"at Augsburg,
Universit\"atsstrasse 1, D-86135 Augsburg.}
\date{\today}

\begin{abstract}
The flow profile in a capillary gap and the pumping efficiency of an
acoustic micropump employing 
Surface Acoustic Waves is investigated both experimentally and theoretically. Such ultrasonic 
surface waves on a piezoelectric substrate strongly couple to a thin liquid layer and generate 
an internal streaming within the fluid. Such acoustic streaming can be used for controlled 
agitation during, e.g., microarray hybridization.  We use fluorescence
correlation spectroscopy and 
fluorescence microscopy as complementary tools to investigate the
resulting flow profile. 
The velocity was found 
to depend on the applied power somewhat weaker than linearly and to
decrease fast with the distance from the 
ultrasound generator on the chip.
\end{abstract}

\pacs{47.85.Np, 83.50.Xa, 83.10.Pp, 43.25.Nm, 47.15.Gf, 87.64.Tt}

\maketitle


\section{INTRODUCTION}
For a biological hybridization experiment, typically, single stranded
DNA is covalently bound to small probe spots on the surface of a glass
slide, forming a so-called micro array. A solution containing fluorescently labelled sample 
molecules is then spread across this micro array where the molecules
are allowed to bind to the respective probe spots.  Successful
hybridization is usually observed as an increase in the fluorescence
intensity on a spot. To have a reference, in co-hybridization
experiments a two color coding is used. Usually, such experiments are 
performed in  sealed chambers to prevent the thin liquid film above 
the slide from evaporation. Given the typical film thickness of less than 100 
microns, the system exhibits a very small Reynolds number, and the
molecules in the fluid can only move diffusively \cite{Kamholz2001}. 
Diffusion, however, is a very slow process. Even for small molecules
such as Rhodamine the distance travelled in water 
after $10^{3}$ s is only about 1mm. For relatively large DNA molecules 
the diffusion driven displacement will be even 
smaller. On the contrary, the binding reaction of complementary DNA 
strands is a relatively fast process. Especially in the case when 
the concentration of probe molecules on a spot is 
much larger than the one of their counterparts in solution, most 
of the molecules close to their binding partners at the glass 
slide will become immobilized after a short time. This can lead to the
formation of a  depletion zone with a lower concentration close to the 
solid surface. To reach the equilibrium concentration of bound DNA, molecules have 
to diffuse through the depletion zone, which becomes the time limiting 
bottleneck for the hybridization process. Typical hybridization assays 
are therefore performed over night or even over weekend, and then are stopped at some 
more or less arbitrary point. Such experiments are also referred to as "end point" experiments.
Apart from the above mentioned non-equilibrium conditions, shorter
hybridisation times also lead to a lower fluorescence 
signal intensities. Quite obviously, this process can be accelerated 
if the molecules were to be actively transported along the solid 
surface plane. Due to the low diffusion constant of large DNA strands 
even a slow movement can speed up the hybridisation considerably. 
Another aspect is that hybridisation is often conducted with more
than one kind of DNA with different fluorescent labels
simultaneously. The fluorescence intensities of the different 
immobilized sample molecules are then evaluated quantitatively 
to interpret the results. However, when the DNA binding 
reactions were not in equilibrium at the end of the hybridisation time,
the results can be misleading.

As the amount of DNA available for a hybridisation experiment is often
very limited one has to reduce the volume as much as
possible. Therefore, pump driven flow systems \cite{Darton2001} 
usually having large dead volumes are not convenient to induce 
the molecule transport during the hybridisation process. There 
are numerous other possibilities to actuate thin fluid films, amongst them for example 
electrokinetic forces \cite{Wu2003} \cite{Zeng2001}. 

In this report, we describe a novel technique to actuate a thin 
liquid film in a capillary gap. We employ surface acoustic waves 
(SAW) propagating on the surface of a piezoelectric solid to induce 
acoustic streaming and mixing within the fluid. To understand the 
fluid dynamics in such a system, we concentrate on how the acoustic streaming 
transforms into a velocity field profile within a homogeneous DNA
solution in the capillary gap. 

A surface acoustic wave (SAW) can be excited by the application of a
radio frequency signal (RF) to an interdigital transducer (IDT) on a 
piezoelectric substrate. The wavelength of the excited SAW is defined 
by the geometry of the IDT. Typical displacement amplitudes of a SAW 
are in the nanometer range, depending on the applied RF power. If 
the SAW supporting surface of the substrate is in contact with 
a liquid, and if the SAW has any displacement component normal to the 
surface, ultrasonic SAW power is leaking into the fluid in form of an 
acoustic wave. Therefore, this so-called leaky 
SAW (LSAW) is expected to decay exponentially with distance from its source. 

The interaction between the SAW and the liquid film leads to an
internal, acoustically induced streaming. The reason for this 
is a net pressure gradient in the direction of the sound propagation in the fluid. 
Usually, this leads to the generation of a fluidic jet, which can be 
nicely visualized employing a dye solution and a fluorescence 
video microscope. The captured images are evaluated in a subsequent
image processing step, being a versatile tool for the interpretation 
of dynamic processes. This method yields real images of the observed 
sample as a whole, and therefore reduces the problem of a false interpretation of the data. 
For more precise measurement of the flow field and velocity, we use fluorescence correlation 
spectroscopy (FCS) \cite{Maiti1997}. In this method, the movement of 
fluorescent particles can be determined in the focal spot of a confocal microscope 
by evaluating the autocorrelation function of the intensity
fluctuations. This represents the rate 
of particles entering and leaving the small observation volume.
\section{The experimental setup and microfluidic characteristics}

We assembled a special microfluidic arrangement to induce streaming in
a narrow capillary gap between a glas slide and a plexiglass block, containing a SAW device.
\begin{figure}
\includegraphics[width=8.4cm]{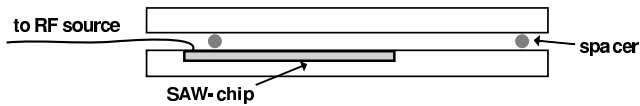}
\caption{The measurement setup for the evolution of dye spots consists 
of a 200 $\mu$m slit between two plexiglas plates (3x3 cm$^2$) kept 
in the right distance with spacers. Integrated in the lower plate is the SAW chip, 
being connected to the RF source outside the fluid film. The
transducer is located in the middle of the plate. 
Images were taken with a video microscope from above through the upper plate.}
\label{fig:setup}
\end{figure} 
 \begin{figure}
\includegraphics[width=8.4cm, angle=180]{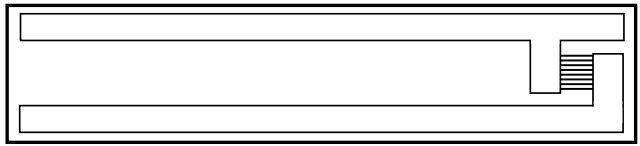}
\caption{The graph shows the sketch of the 4x23 mm$^2$ LiNbO$_3$ chip 
with gold structures. Situated on the right side is the interdigitated 
transducer(1x1 mm$^2$) inducing the SAW perpendicular to the long 
side of the chip. The contact wires are connected with the Rf power 
source on the left end.}
\label{fig:chip}
\end{figure}
Fig.~\ref{fig:setup} gives a sketch of the principal setup as seen
from the side. Fig.~\ref{fig:chip} shows the chip architecture.
The width of the gap is controlled by spacers, it can be continuously
adjusted down to 
30$\mu$m. In the measurements described, we worked with a spacers of 200 $\mu$m.
The SAW chip consists of a 128$^o$ rotated Y-cut, X propagating LiNbO$_3$ single crystal chip. 

The IDT consists of 22 finger pairs, 0.85 mm aperture, and a period
$\lambda = 28 \mu$m. The IDT fingers are thermally evaporated 300 nm 
thick Au electrodes, fabricated by a standard photolithografic technique.
The complete chip was protected with a 800 nm thick, sputtered Silicon 
Dioxide layer, which was removed at the contact pads to allow for 
electrical connection.
The IDT was driven with an AC signal at the resonance frequency $f_0$, being defined as 
$f_0 =c/\lambda$.
For a SAW velocity $c \approx $ 3800 m/s, the
frequency  $f_0$ is hence about 136 MHz.
Such an IDT efficiently converts the applied RF signal into an
acoustic wave, which in this case is launched as a bidirectional 
beam perpendicular to the fingers.

\begin{figure}
\includegraphics[width=8.4cm]{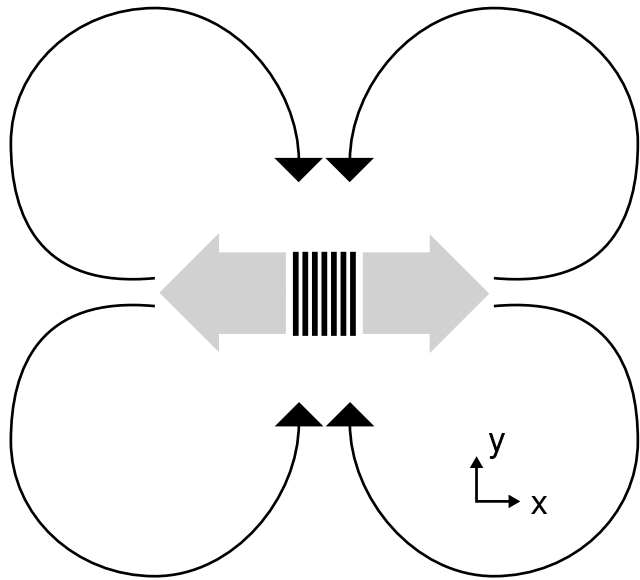}
\caption{Flow pattern generated by SAW. The fluid is 
pumped perpendicular to the metal fingers of the IDT and is sucked in
parallel to them.}
\label{fig:stromlin}
\end{figure}

If the capillary gap is filled with a fluid, switching on the RF 
power leads to a flow pattern as sketched in Fig.~\ref{fig:stromlin}.
The fluid is expelled from the transducer region to the left and 
right side and flows back to the IDT from bottom and top.
According to the high viscous damping the flow velocity decreases 
rapidly with distance from the IDT.

Further investigations were focused on the characterization of the
flow field. The issue of this work was to find out how the setup can 
be optimized for agitating a large hybridisation area.
For this purpose 
we first want to discuss the nature of the observed SAW-induced motion.
\par
The propagating LSAW couples to a sound wave in the liquid.
Because the sound velocity in fluids $c_{\text{liquid}}$ is always smaller than
the LSAW-velocity in the solid substrate, 
sound is radiated into the liquid under an angle obeying Snellius' law of diffraction:
\begin{equation}
\label{eq:SnellsLaw} \sin( \theta_R) = \frac{c_{\text{liquid}}}{c_{\text{LSAW}}}
\end{equation}
This can be understood as a consequence of phase matching between the 
LSAW and the radiated sound beam in the liquid.
This "mode conversion" leads to an exponential decay of the LSAW. 
In our case the observed damping in power was about 1dBm per
wavelength, which is in good accordance with values found in refs. 
\cite{Shiokawa1989} and \cite{Campbell1970}.  
This results in a decrease of 35dBm per mm. In the following we will 
give a short overview how the sound radiated in the fluid can induce a streaming force.

Sound travelling through a liquid is attenuated by the viscosity 
along its transmission through the medium. If the intensity is 
high, this attenuation creates an acoustic pressure gradient along the propagation 
of the wave. The gradient induces a force in the same direction that
causes a flow in the fluid. This conversion of an attenuated sound
wave into a steady flow is a nonlinear effect which is known as acoustic streaming.    
When a SAW, i.e. a surface acoustic wave with a spatially constant amplitude, 
is in contact with a fluid the acoustic streaming is only
a minor effect \cite{Shiokawa1989}. Much stronger is the streaming 
that results from the exponential decay of the LSAW.
This decay causes the conversion into sound waves in the liquid within a
short distance leading to a large gradient of the sound
amplitude and consequently to a strong force that generates the acoustic streaming.

Here we do not want to go into the details of the rather complex
mechanism of acoustic streaming \cite{Nyborg1965}, \cite{Uchida1995}. 
We only mention that the effective
force that causes the steady flow can be expressed in terms of the
sound velocity field in the fluid \cite{Nyborg1965}:
\begin{equation}
\label{eq:Streaming Force} {\bf F}= 
\rho \langle  ({\bf v}_f\cdot \nabla){\bf v}_f +{\bf v}_f(\nabla\cdot {\bf v}_f) \rangle
\end{equation}
where the angular brackets denote time average that removes the fast
oscilations of the sound wave.

Finally we note that the force is proportional to $\omega^2$. High
force densities therefore can be achieved by means of high LSAW
frequencies. On the other hand, lower frequencies lead to smaller
damping rates of the LSAW and the force density acts on a larger
volume. For an optimal design a trade off has to be found.


\section{MATERIALS AND METHODS}

\subsection{Fluorescence Microscopy}
\paragraph{Microscope}
To visualize the activity of the SAW 
transducer, a Na-fluorecein solution (1mg/ml) is pipetted into a small hole in the cover
of the fluid film. 
For the excitation of the fluorescence of the dye, blue LEDs are 
attached to a ring some distance above the cover. Fluorescence images are taken by a video microscope (Olympus, Hamburg) with a suitable filter to suppress the 
excitation light.

\begin{figure}
\includegraphics[height=8.4cm,angle=-90]{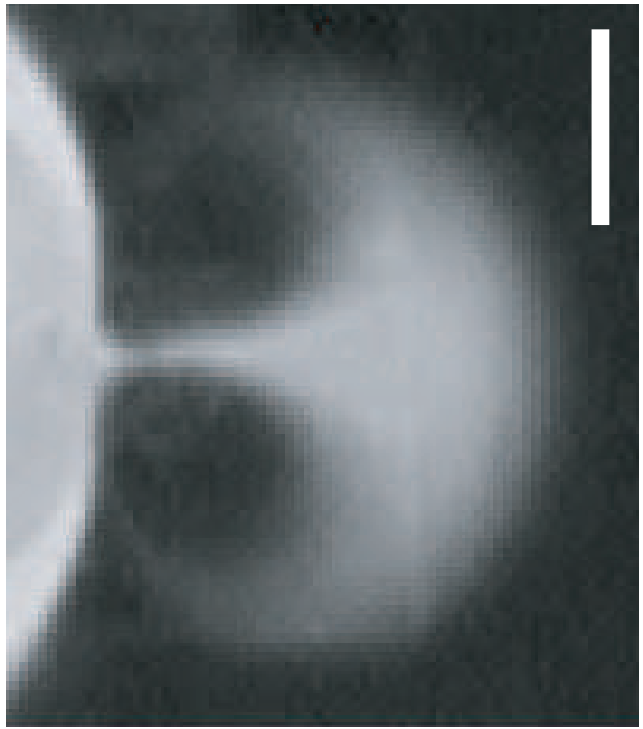}
\caption{A dye jet pumped by the SAW transducer, shortly after switching the RF-power.
In the middle of the transducer the fluid is pumped upwards, 
on both sides the fluid is sucked in.
The bar is 0.5 mm wide.}
\label{fig:jet2}
\end{figure}

\begin{figure}
\includegraphics[width=8.4cm]{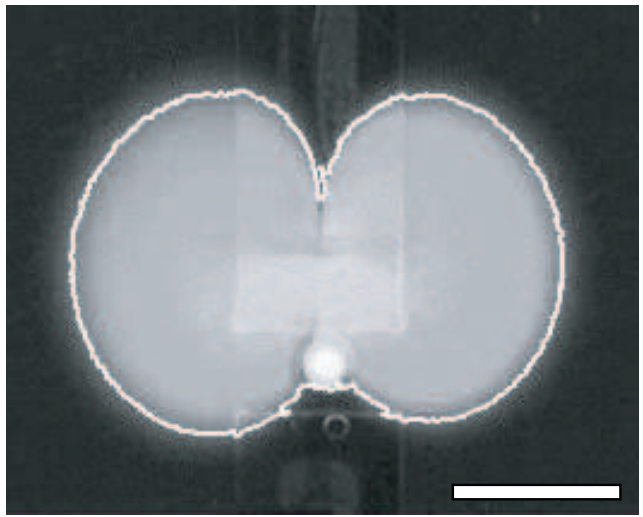}
\caption{A dye spot after 5 min of SAW pumping (40 mW) is shown. 
Superimposed is the edge line calculated by 
the image procedure. The bar is 1mm wide}
\label{fig:umrand_s}
\end{figure}

\begin{figure}
\includegraphics[width=8.4cm]{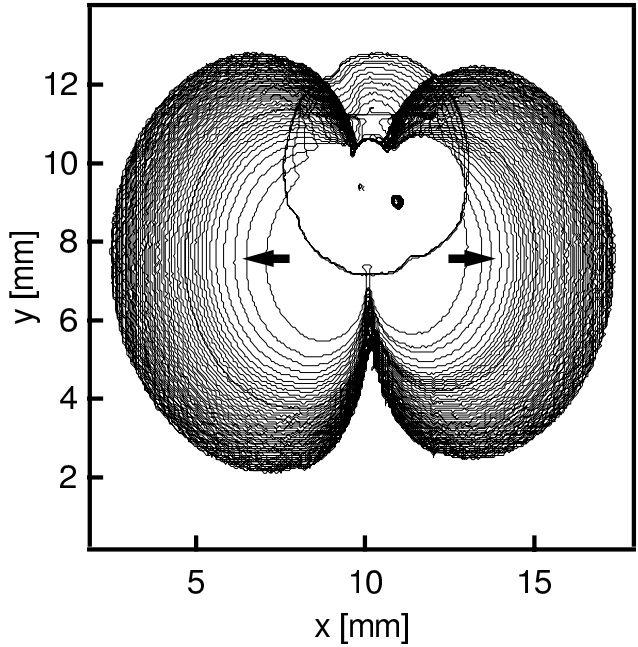}
\caption{The edges of a dye spot during the pumping of the SAW transducer (40 mW) are recorded
every 10 s and put together in one graph. 
The evolution from a round spot to a butterfly shape can be
observed.}
\label{fig:schmetterl2}
\end{figure}

\begin{figure}
\includegraphics[width=8.4cm]{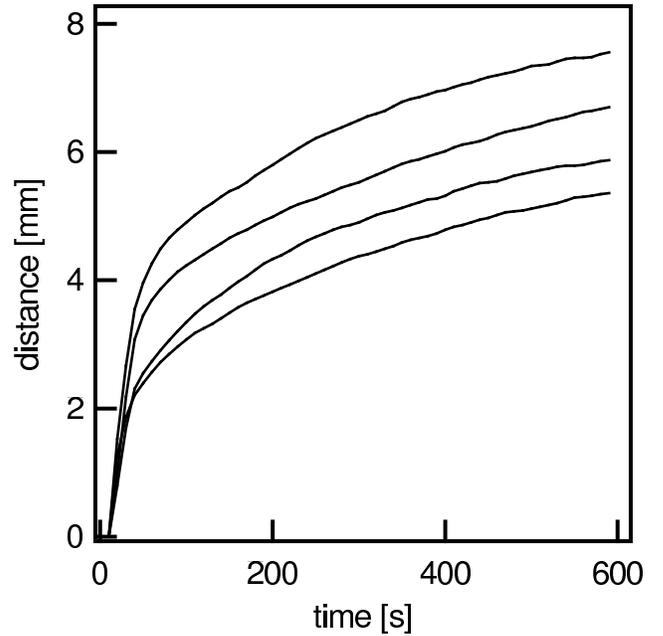}
\caption{The distance of the dye front as a function of time for 
10, 25, 63 and 158 mW RF-power (from bottom to top). 
The data points where determined every 10 seconds at the line, 
marked by the arrows in Fig.~\ref{fig:schmetterl2} and connected by lines.}
\label{fig:distance}
\end{figure}

\paragraph{Image processing}
Fluorescence images of the evolving Na-fluorescein spot
(see Figs.~\ref{fig:jet2}, \ref{fig:umrand_s} were taken 
every 10s and stored into an avi movie file. The avi movie was 
further processed by a procedure converting the pictures 
into binary images with the help of the Shannon entropy function 
and detecting the edges of the dye spot with the particle finder 
algorithm. The coordinates of every edge line were put together 
in one file (Fig.~\ref{fig:schmetterl2})
and the positions of the leading edge to the left 
and the right was extracted for every recorded point of time. 
This leads to a distance vs time diagram (Fig.~\ref{fig:distance})
. 
Differentiating this data yields a speed vs distance diagram. 
Also, the area of the Na-fluorescein spot and the speed of the area 
increase can be calculated.

\subsection{Fluorescence Correlation Spectroscopy}

In 1972 Magde, Elson and Webb published the first work on FCS \cite{Magde1972}.
In the following years the same authors elaborated the technique and it's
applications in detail \cite{Elson1974} \cite{Magde1974} \cite{Magde1978},
where in the last reference theory and first experiments with FCS on systems
with laminar flow are presented. Until the early 1990s FCS was mainly used to
determine diffusion properties and kinetic constants of chemical reactions in solution. In 1993 the introduction of a confocal detection optics by Rigler et
al. \cite{Rigler1992} gave new impulses to the method. Schwille
\cite{Schwille1997a}, Krichevsky \cite{Krichevsky2002} and Thompson
\cite{Thompson2002} have given reviews about the principles, applications and
potentials of FCS.

The raw signal in FCS experiments is the time-dependent intensity signal coming
from a small and fixed volume ($\approx 0.5 \unit{fl}$) created by a laser
focus. Fluorescently labelled molecules entering and leaving the open volume
lead to fluctuations in the recorded signal $F(t)$:
\begin{equation}
    \label{eq:intensity} F(t) = q \, Q \, \int_{V} I(\vec{r}) C(\vec{r},t) dV
\end{equation}
Here, $q$ stands for constant intensity losses due to the experimental setup
e.g. detector efficiency, filters. $Q$ is the fluorescence quantum yield of the
dye and $C(\vec{r,t})$ denotes the concentration of the labelled particles,
$I(\vec{r}) = I_0 \exp(-2 (x^2+y^2)/r_0) \exp(-2 z^2/z_0)$ is a
combination of the collection efficiency and the excitation profile. If the
confocal detection geometry is chosen properly a Gaussian effective volume with
radius $r_0$ and height $z_0$ is a good approximation \cite{Hess2002}.

In this work we want to focus on the possibility to measure flow velocities as
it was shown by G\"osch et al. \cite{Goesch2000}.

The experimental autocorrelation function $g_2(t)$ is calculated from the
measured fluorescence intensity $F(t)$ by
\begin{equation}
    \label{eq:g2} g_2(t) = \frac{\langle F(t_0) F(t_0+t) \rangle}{\langle F(t_0) \rangle
    ^2}= 1 + \frac{\langle \delta F(t_0) \delta F(t_0+t) \rangle}{\langle F(t_0) \rangle
    ^2}.
\end{equation}
For Brownian diffusing particles the concentration $C(\vec{r},t)$ is known and
can be used to calculate an expected autocorrelation function by combining
equation (\ref{eq:g2}) and equation (\ref{eq:intensity})
\begin{equation}
    \label{eq:g2Diff} g_2(t) = 1 + \frac{1}{N} \, G\ind{Diff}(t)
\end{equation}
with $N$ denoting the average number of marked objects in the effective focal
volume and
\begin{equation}
    \label{eq:GDiff} G\ind{Diff}(t) = \frac{1}{1 +
      \frac{t}{\tau\indt{D}}} \, 
\frac{1}{\sqrt{1 + \frac{t}{f^2 \tau\indt{D}}}}
\end{equation}
where $\tau\ind{D} = \frac{r_0^2}{4 \, D}$ denotes the mean passage
time of the particle 
through the effective volume \cite{Eigen1994}.
and $f = z_0/r_0$ gives the aspect ratio of the effective volume. With the
Rhodamine 6G diffusion constant $D\ind{R6G} = 280 \, \mu\mbox{m}^2
\mbox{s}^{-1}$~\cite{Magde1974} as a reference, the focal waist size and shape
parameter were determined ($r_0 \simeq 0.20\,\mu$m and $f \simeq 8$), and used
for all subsequently acquired data.

Equation (\ref{eq:GDiff}) turns out to be the solution for the ideal case of
point-like particles, which is well describing the case where the particle size
$R$ is small compared to the focal volume size $r_0$. For situations where $R$
is comparable to, or even larger than $r_0$ no analytically closed form of equation
(\ref{eq:GDiff}) can be given (particle size effect). The autocorrelation
function depends strongly on the distribution of fluorophores on the particle.
For spherical particles with radius $R$ an approximate diffusion time can be
given \cite{Starchev1998}.
\begin{equation}
    \label{eq:tauDPartSize} \tau\ind{D} = \frac{r_0^2 + R^2}{4 \, D}
\end{equation}
In the presence of a uniform translation in addition to the diffusive
particle motion the autocorrelation function becomes 
\begin{equation}
    \label{eq:g2DiffFlow} g_2(t) = 1 + \frac{1}{N} \, G\ind{Diff}(t) \, G\ind{Flow}(t)
\end{equation}
where
\begin{equation}
    \label{eq:GFlow} G\ind{Flow}(t) = \exp (- \frac{t}{\tau \indt{F}})
\end{equation}
Taking into account the particle size effect the dwell time $\tau \indt{F}$ of
the particles in the focus is given by
\begin{equation}
    \label{eq:tauFPartSize} \tau\ind{F} = \frac{r_0 + R}{v \indt{Flow}}
\end{equation}
As can be seen from equation (\ref{eq:g2DiffFlow}) the shape and the
characteristic decay time of the autocorrelation function is dominated by the
faster of the two competing processes. In order to obtain the flow
velocities without perturbation by diffusion we used fluorescent
microspheres of 1$\mu m$ radius (Fluosperes Yellow-Green 505/515,
F-8823, Molecular Probes, Eugene, OR). On the experimentally relevant
time-scale the Brownian motion of these
spheres is of minor importance. Therefore we
could neglect the diffusive term in equation (\ref{eq:g2DiffFlow}).

We used a commercial setup by Carl~Zeiss (Jena, Germany) consisting of the
module Confocor~2 and the microscope model Axiovert~200 with a Zeiss
C-Apochromat 40x NA 1.2 water immersion objective. For excitation, the
$514\,$nm line of a $25\,$mW Argon laser was used. Emitted fluorescence was
detected at wavelengths longer than $530\,$nm by means of an avalanche photo
diode allowing for single-photon counting.

\section{RESULTS}

As a test for the image processing, a spot of Fluorescein pipetted 
into the hole of the cover plate was observed for 10 min., with 
the SAW switched off. The area covered by the dye therefore only
increases by diffusion. Using the linear diffusion equation
$\langle x^2\rangle =2Dt$, 
a value for the diffusion constant of the dye molecules 
of D=370$\mu\unit{m}^2/\unit{s}$ can be determined which is in good agreement
with  literature values of 200-400$\mu\unit{m}^2/\unit{s}$.

\begin{figure}
\includegraphics[width=8.4cm]{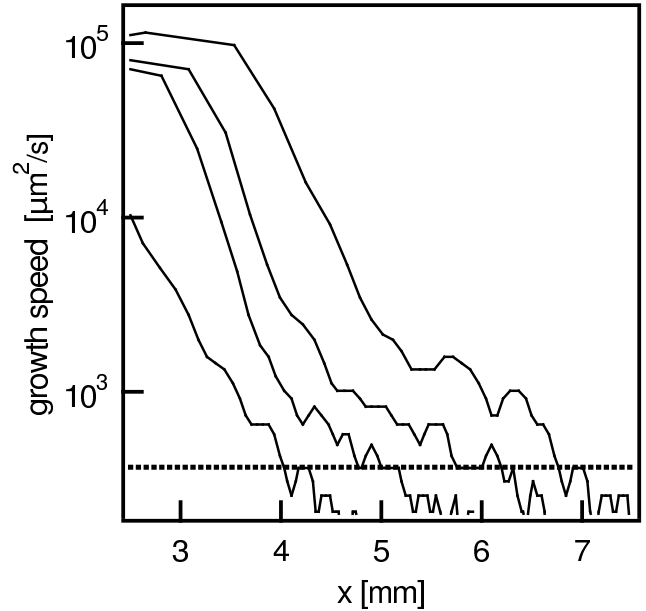}
\caption{The speed of the growth of the area of the 
Fluorescein spot is plotted against the particular
position of the advancing dye front for 4 different RF-power values 
(10, 40, 63 and 158 mW from left to right). 
When the speed reaches the region of the diffusion constant of the dye,
the data is superposed by noise. The diffusion constant of
Fluorescein (370 $\mu\unit{m}^2/\unit{s}$) is marked by a dashed line.
For small distances, representing early times, the image processing yields 
misleading results, since the transducer region is covered by
the initial round dye spot.}
\label{fig:4_area}
\end{figure}

\begin{figure}
\includegraphics[width=8.4cm]{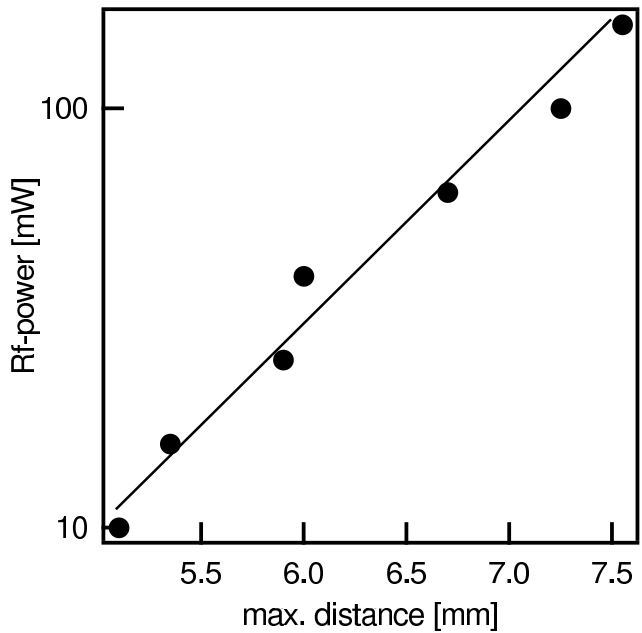}
\caption{The distance $ x_D^*$  where the speed 
of the area growth reaches the diffusion constant of
Fluorescein (370 $\mu\unit{m}^2/\unit{s}$) dependening on the applied RF-power. Superimposed
to the data is a straight line.}
\label{fig:flleist}
\end{figure}

\begin{figure}
\includegraphics[width=8.4cm]{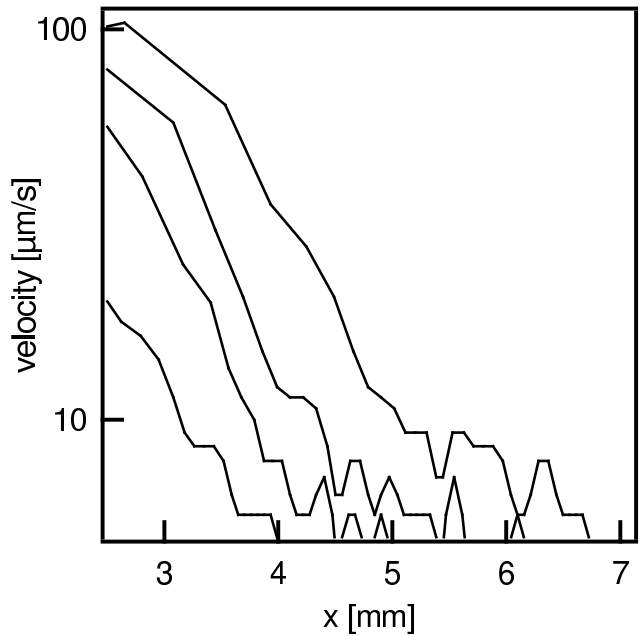}
\caption{The speed of the propagating dye front in the direction of
  the arrows in Fig.~\ref{fig:schmetterl2}  
is plotted against the particular position for 4 different RF-power 
values (10, 40, 63 and 158 mW from left to right).
 The speed values are calculated by
differentiating the front position vs time data. Small distances are
not displayed, because the image 
processing yields misleading results close to the transducer.}
\label{fig:4_speed}
\end{figure}

\begin{figure}
\includegraphics[width=8.4cm]{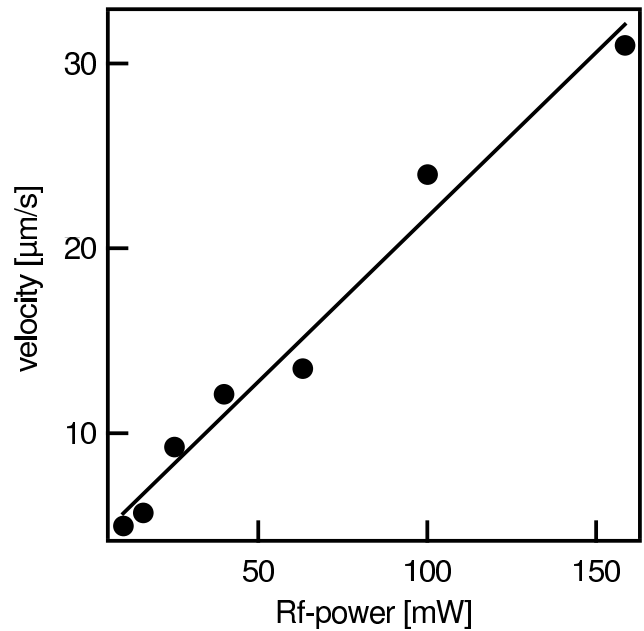}
\caption{The speed of the advancing dye front at a distance 4mm from the SAW transducer
is plotted against the RF-power. A straight line has been fitted to the data.}
\label{fig:4mm_2}
\end{figure}

In Fig.~\ref{fig:jet2} a close up image of the area near the transducer 
is shown, shortly after starting the SAW. The 1mm wide IDT is hidden 
by the dye solution on the lower side pumping a 200 $\mu$m wide jet 
to the upper side. The jet becomes broader with increasing distance 
and a part is transported to the influx areas of the IDT 
perpendicular to the direction of pumping. After 10 min., 
the Fluorescein covered area shows a butterfly shape. 
Employing the image processing the edge of the dye spot is determined
every 10 s during the observation time.
A border line between the colored and uncolored areas computed this 
way is plotted in Fig.~\ref{fig:umrand_s}. Collecting the traces 
of a 10 min. experiment into one graph 
yields a plot as in Fig.~\ref{fig:schmetterl2}. The initial dye 
spot and the 
succeeding evolution of the edge lines are visible in this graph. This data 
was used for determining the evolution speed of the area (Fig.~\ref{fig:4_area}).
The growth of a Fluorescein spot was observed for 7 different 
RF-powers and the actual position of the front line in the direction 
of the arrows in Fig.~\ref{fig:schmetterl2} was recorded. The growth speed 
decreases fast with distance from the transducer, until the diffusion constant of the 
dye (370 $\mu\unit{m}^2/\unit{s}$) is reached. 
In the experiment, this position $ x_D^*$ is found to depend linearly
on the logarithm of the RF-power, see Fig.~\ref{fig:flleist}.  

From the movement of the dye front in the symmetry direction marked by the arrows 
in Fig.~\ref{fig:schmetterl2}
the position can be determined as a function of time (Fig.~\ref{fig:distance}). 
Differentiating this data and plotting it against 
the x-position of the front leads to a fast decrease of the speed
(Fig.~\ref{fig:4_speed}), 
similar as found for 
the growth of the dye covered area. The decay of the velocity with the
distance does not change qualitatively upon varying the RF-power.
As a quantitative measure we show the velocity 
at the 
distance 4mm from the transducer in dependence of the applied
RF-power in Fig.~\ref{fig:4mm_2} . In the measured range a linear dependence yields a
reasonable fit although the extrapolation to small powers becomes unphysical. 

\begin{figure}
\includegraphics[width=8.4cm,clip]{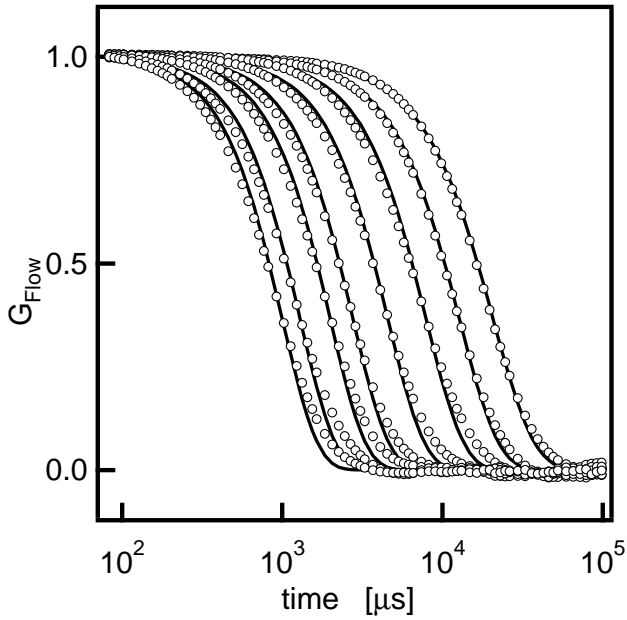}
\caption{Normalized autocorrelation curves from FCS measurements. The applied
power was varied between $150 \unit{mW}$ (right) and $500 \unit{mW}$
(left) in steps of $50 \unit{mW}$.
Circles represent the experimental data whereas the lines are fits to equation
(\ref{eq:GFlow}). } \label{fig:g2}
\end{figure}

\begin{figure}  
\includegraphics[width=8.4cm]{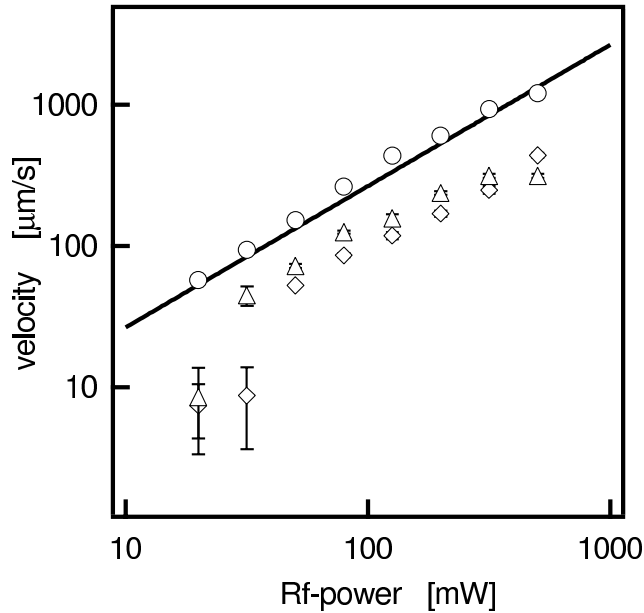}
\caption{ Flow velocity as a function of the applied power for three different
positions indicated by circles, triangles and squares. 
The corresponding positions are marked by the same
symbols in the Fig.\ref{fig:velpos}. 
Circles represent the results from the fitted curves in Fig.
\ref{fig:g2}. The line is a fit to the data with a slope of $1$. Flow
velocities close to 10 $\mu m/s$ are not determined correctly (points
with large error bars).}
\label{fig:velpow}
\end{figure}
\par
Fig.~\ref{fig:g2} shows FCS correlation data taken as a function of the applied
power. The correlation coefficient $G_{\text{Flow}}(t)$ was
determined from the measured intensity autocorrelation by means of the
eqs. (\ref{eq:g2}) and (\ref{eq:GFlow}) under the assumption that
$G_{\text{Diff}}(t) = 1$, and compared to the
exponential law (\ref{eq:GFlow}). The exponential fits well agree with
the experimental data. 
The small deviations 
may be attributed to particle size effects \cite{ Starchev1998} 
or to the polydispersity of the used beads
\cite{ Starchev1999}. 
The fitted dwell times of the particles
show a strong dependence on the applied power. According to equation
(\ref{eq:tauFPartSize}) the velocity was calculated and plotted as a function
of the power (Fig.~\ref{fig:velpow}). This was done for three different
positions in the channel. The highest velocities were achieved at positions
near the chip. The measured data can  reasonably well be fitted by a
linear dependence of the velocity on the applied power. 

\begin{figure}
\includegraphics[width=8.4cm,clip]{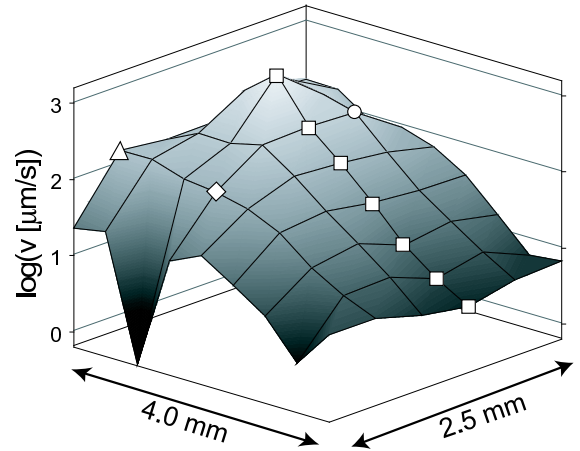}
\caption{ Velocity profile for a selected area near the pumping
  chip at a RF-power of 79 mW. The triangle, square and circle mark the positions at which
  the velocities shown in Fig.~\ref{fig:velpow} were taken.}
\label{fig:velpos}
\end{figure}
\begin{figure}
\includegraphics[width=8.4cm,clip]{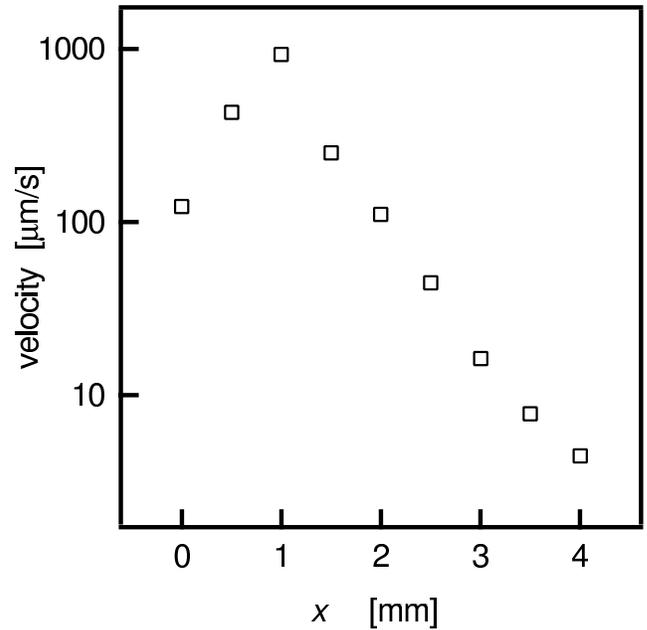}
\caption{ Velocity along the central x-direction at the positions
  shown by squares in Fig.~\ref{fig:velpos}. Ideally, the velocity at
  the symmetry point $x=0$ should vanish. The high gradient of the
  velocity field there and the experimental positional uncertainty, however,
  lead to the apparent finite velocity.} \label{fig:velxpos}
\end{figure}

In Fig.~\ref{fig:velpos} the velocity profile for a selected area close to the chip
is shown. The $x$-direction is the direction in which the chip is supposed to
pump. To investigate the dependence quantitatively we took the values indicated
by the open squares and plotted their logarithm as a function of the
position, see
Fig.~\ref{fig:velxpos}. As can be seen from this graph, the velocity decreases
in a similar way as found for the propagation of the dye front in Fig.~\ref{fig:4_speed}.

\section{Theory}
\subsection{Calculation of the flow field}
All observed flows in the 200$\mu$m thick fluid film 
had  velocities smaller than 1 mm/s and, hence, are characterized  by
Reynolds numbers smaller than 1.   
This means that the flow 
is completely laminar \cite{Brody1996}, which simplifies the situation for 
the calculation of the streamlines.   
Although the detailed mechanism of the sound conversion into steady
streaming is fairly complicated the steady flow is determined by the
relatively simple Stokes equations \cite{Nyborg1965} : 
\begin{equation}
\label{eq:Stokesflow} \mu \Delta {\bf v}({\bf x}) = \nabla {p}({\bf
  x}) -{\bf F}( {\bf x})
\end{equation} 
\begin{equation}
\label{eq:incompressible} \nabla \cdot {\bf v} = 0 
\end{equation} 
where $\mu$ denotes the shear viscosity of the fluid.
The driving force ${\bf F}$ is given in terms of the sound wave by
eq. (\ref{eq:Streaming Force}). 
The SAW force field is generated in the area of the transducer and
decays exponentially with distance. At about 1 mm it has dropped 
more than three orders of magnitude. 
We use these features in order to find an approximate analytical solution of
eqs. (\ref{eq:Stokesflow}) and (\ref{eq:incompressible}) for the time
averaged velocity and pressure fields and simplify the problem
by the following assumptions: (i) lateral boundaries are neglected,
i.e. we consider an infinitely extended fluid layer of thickness $d$
with no-slip boundary conditions at both confining plates;
(ii) we approximate the actual spatially extended force density by
a pair of constant forces being concentrated on lines perpendicular
to the plates separated by the distance $a$: 
\begin{equation}
{\bf F}({\bf x}) = \left ( {\bf f}_+ \delta(x-a/2) + {\bf f}_-
\delta(x+a/2) \right ) \delta(y)
\label{FD}
\end{equation}
where $(x,y) ={\bf r} $ are coordinates  spanning the planes parallel to the
plates and $z$ is a coordinate perpendicular to the plates with $z=0$
on lower and $z=d$ on the upper plate. 
The two forces ${\bf f}_{\pm}$ are assumed to
have the same component in the $z$-direction and 
opposite components in the direction of the vector separating the two
forces, i.e. in the $x$-direction, pushing the fluid away, 
\begin{equation}
{\bf f}_\pm = f_z {\bf e}_z \pm  f {\bf e}_x.
\label{f}
\end{equation}
where ${\bf e}_{x,z}$ are the unit vectors in $x$- and $z$-direction, respectively.  
We expect that outside the narrow spatial domain where the actual
force does not vanish this presents a fairly good approximation to the
actual situation.
Because of the linearity of the Stokes equations
(\ref{eq:Stokesflow},\ref{eq:incompressible}) it is sufficient to
consider a single line-force contribution. The final result can be
obtained by the superposition of the solutions to the two forces. 
      
\begin{figure}
\includegraphics[width=7.cm]{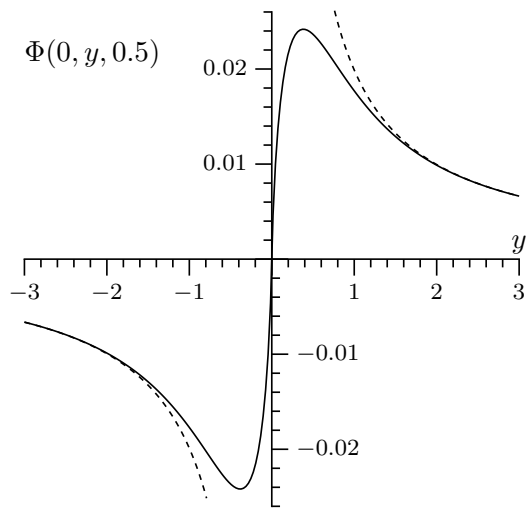}
\caption{The streaming function $\Phi(0,y,0.5)$ (solid line)
  perpendicular to the
axis $x=0$ displays extrema close to $|y|= 0.3$. The broken line
shows the asymptotic
far field behavior according to eq.~(\ref{Phias}). Both functions perfectly coincide
for distances from the origin that are larger than twice the
plate distance. }
\label{phi}
\end{figure}

\begin{figure}
\includegraphics[width=7.cm]{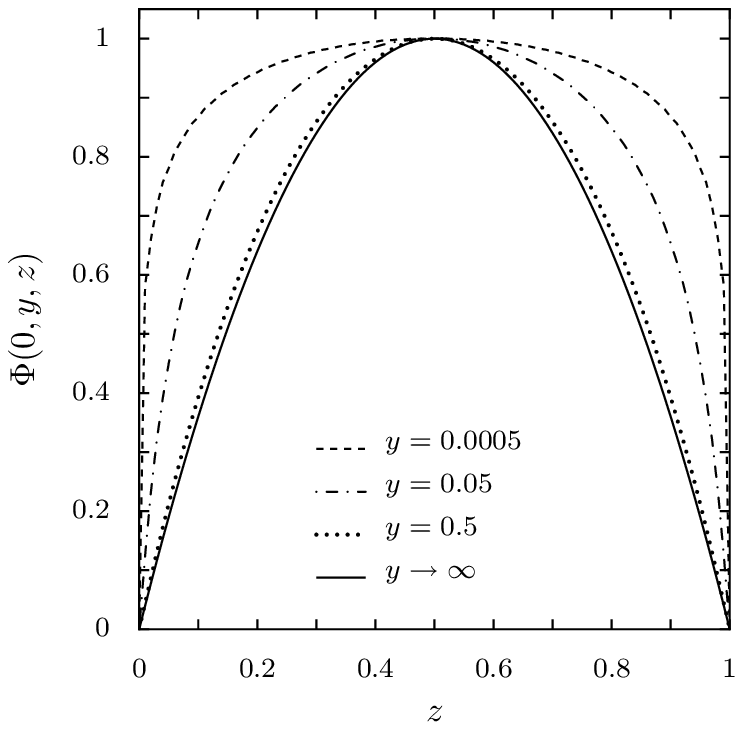}
\caption{The streaming function $\Phi(0,y,z)/\Phi(0,y,0.5)$ divided by its maximal value at
  $z=0.5$ at $x=0$ is shown as a function of the transversal
  coordinate $z$ for different distances $y$ from the
  origin where the force is sitting. In the very close vicinity  
  of the force source, this profile is
  flat in the middle part of the fluid layer. For larger
  distances from the force it rapidly approaches the asymptotic
  parabolic profile, see eq.~(\ref{Phias}).}
\label{FP}
\end{figure}

\begin{figure}
\includegraphics[width=8.4cm]{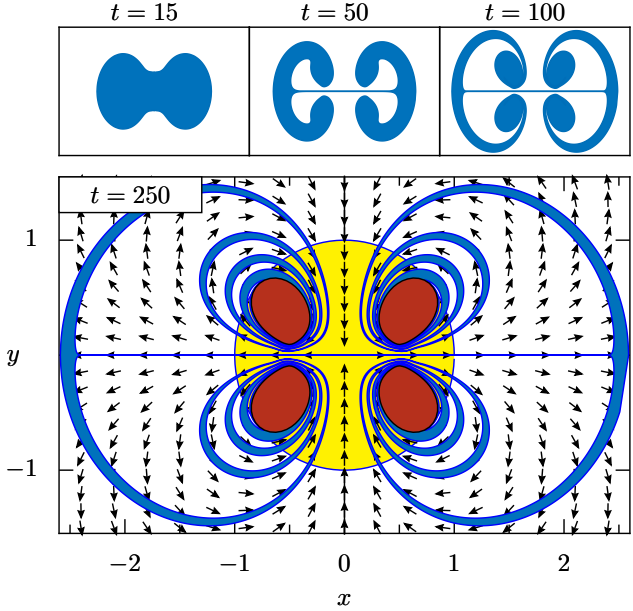}
\caption{The graph shows the calculated evolution of a dye spot pumped by a SAW transducer 
in the center of the images. The black arrows mark the 
direction of the velocity field. The initial dye spot is marked by the circle. 
The 4 black ellipses in the lower image represent areas where the dye is trapped
by an eddy. The trajectories were calculated for the plane in the middle
of the fluid film with diffusion neglected.}
\label{fig:flow}
\end{figure}

Having only one line-force we may put the position of coordinate center at
the position of the force. Further we note that one formally may include the
$z$-component of the force into the pressure term by introducing the
effective pressure
\begin{equation}
p_{\text{eff}}({\bf x}) = \tilde{p}_2({\bf x}) - f_z z \delta(x) \delta(y)
\end{equation}
Measuring lengths, velocities and pressure in units of $d$, $f/\mu$
and $f/d$, respectively, leads to the dimensionless Stokes equations:
\begin{eqnarray}
\Delta {\bf v}({\bf x}) - \nabla p_{\text{eff}} ({\bf x})& = &- {\bf e}_x
\delta^2 ({\bf r})\\
\nabla \cdot {\bf v}({\bf x}) = 0
\label{Stdl}   
\end{eqnarray}
with no-slip boundary conditions on the plates:
\begin{equation}
{\bf v}({\bf r},0) ={\bf v}({\bf r},1) =0
\label{bc}
\end{equation}
where $\delta^2({\bf r}) = \delta(x) \delta(y)$ is a two-dimensional $\delta$-function.
If one takes the divergence of the Stokes equation one finds a Poisson
equation for the effective pressure:
\begin{equation}
\Delta p_{\text{eff}} ({\bf x}) = \frac{\partial}{\partial x} \delta^2({\bf r}) 
\label{Pp}
\end{equation}
Because the inhomogeneity on the right hand side is $z$-independent it
follows that the effective pressure also does not depend on $z$. Consequently
the $z$-component of the velocity fulfills the Laplace equation with
homogeneous boundary conditions and therefore $v_z({\bf x})$ vanishes
everywhere.
The remaining two components of the velocity field can be expressed in
terms of a stream function $\Phi({\bf x})$:
\begin{equation}
v_x({\bf x}) = \frac{\partial \Phi({\bf x})}{\partial y}, 
\quad v_y({\bf x}) = -\frac{\partial \Phi({\bf x})}{\partial x}
\label{vPhi}
\end{equation}
Differentiating the $x$- and $y$-components of the Stokes equation
with respect to $y$ and $x$, respectively, and subtracting them one
obtains for the stream function:
\begin{equation}
\Delta_{\bf r} \Delta \Phi({\bf x}) = - \frac{\partial}{\partial y}
    \delta^2({\bf r})  
 \label{DDPhi}
\end{equation}
where $\Delta_{\bf r} = \partial^2 / \partial x^2 +\partial^2 /
\partial y^2$ denotes the two-dimensional Laplace operator in the
$(x,y)$-plane.
Boundary conditions for $\Phi({\bf x})$ are:
\begin{eqnarray}
\Phi({\bf r},0) & =& \Phi({\bf r},1)  = 0, \\ 
\frac{\partial \Phi({\bf x})}{\partial x} & = & \frac{\partial \Phi({\bf
    x})}{\partial y} =0\\
\lim_{r \to \infty} \Phi({\bf r},z) & = & 0
\label{bcPhi}
\end{eqnarray}
where $r = (x^2 + y^2)^{1/2}$ denotes the absolute value of ${\bf r}$.
Because of the periodic boundary conditions with respect to $z$ the
stream function can be represented as a $\sin$-series:
\begin{equation}
\Phi({\bf x}) = \sum_{n=1}^\infty c_n({\bf r}) \sin (\pi n z)
\label{Phis}
\end{equation}
Here the coefficients are still functions of ${\bf r}$ and obey the
equations
\begin{equation}
\Delta_{\bf r} \left ( \Delta_{\bf r} - (\pi n)^2 \right ) c_n({\bf r})
= -\gamma_n \frac{\partial}{\partial y} \delta({\bf r}) 
\label{cn}
\end{equation}
where 
\begin{equation}
\gamma_n = 2 \int_0^1 dz\: \sin (\pi n z) = \left \{
\begin{array}{cl}
0 \qquad & \mbox{for $n$ even}\\
\frac{4}{\pi n} \qquad & \mbox{for $n$ odd}.
\end{array}
\right .
\label{gamma}
\end{equation}
The solutions of these equations can be expressed in terms of the
Green's functions of the time-independent two-dimensional Klein-Gordon 
equation:
\begin{equation}
\left ( \Delta_{\bf r} - k^2 \right ) g_m({\bf r}) = \delta^2({\bf r})
\label{KG}
\end{equation}
where $k$ is a positive number.
This Green's function  is given by \cite{Albeverio}:
\begin{equation}
g_k({\bf r}) = -\frac{1}{2 \pi} K_0(k r)
\label{gm}
\end{equation}
where $K_\nu(z)$ is the modified Bessel function of order $\nu$, see Ref.~\cite{AS}.
To achieve this goal we introduce an auxiliary function and rewrite
eq.~(\ref{cn}) in terms of two Klein-Gordon equations:
\begin{eqnarray}
\Delta_{\bf r} h({\bf r}) & = & -\frac{\partial}{\partial
  y}\delta^2({\bf r}) \\ 
\left ( \Delta_{\bf r} - (n \pi)^2) c_n({\bf r} \right ) & = & \gamma_n h({\bf r})
\end{eqnarray}
The solution to the first equation can readily be expressed in terms
of the Green's function for $m=0$:
\begin{eqnarray}
h({\bf r}) & = & - \lim_{m \to \infty} \frac{\partial}{\partial y}
g_m({\bf r}) \nonumber \\
& = & \frac{1}{2 \pi} \frac{\partial}{\partial y} \ln r
\label{g0}
\end{eqnarray}
From the second equation one then finds:
\begin{eqnarray}
c_n({\bf r}) & = & \gamma_n \int d^2 {\bf r}' \: g_{n \pi} ( {\bf r}- {\bf
  r}') h({\bf r}') \nonumber \\
& = & \frac{\gamma_n}{(2 \pi)^2} \frac{\partial}{\partial y} \int d^2
  {\bf r}' \: K_0(n \pi r') \ln|{\bf r} -{\bf r}'|
\label{cni}
\end{eqnarray}
Introducing polar coordinates one may perform the integration over
${\bf r}$  and
obtains:
\begin{equation}
c_n({\bf r}) = \left \{ 
\begin{array}{ll}
0 &\text{for $n$ even}\\ 
\frac{2 y}{n^3 \pi^4 r^2} \left (1-n \pi r K_1(n \pi r)
\right )\;&\text{for $n$ odd} 
\end{array}
\right .
 \label{cnf}
\end{equation}
The modified Bessel function decays rapidly for large values of its
argument and therefore $n \pi r K_1(n \pi r)$ can be neglected for $\pi
r \gg 1$. This gives the asymptotic behavior of the stream function:
\begin{equation}
\Phi_{\text{as}}({\bf x})= \frac{y}{4 \pi r^2}z(1-z)
\label{Phias}
\end{equation}
where we summed the sine-series $\sum_{n\; \text{odd}} \frac{\sin (n \pi)
z}{n^3} = \frac{\pi^3}{8}z(1-z)$, see Ref. \cite{BS}. Deviations
from this asymptotic result can only be seen within a distance of a
few gap widths from the location where the force is acting, see
Figs.~(\ref{phi}) and (\ref{FP}). 
The streaming function of the force dipole readily
follows as the linear combination of two velocity fields that are
caused by slightly displaced line-forces pointing in opposite lateral 
directions:
\begin{equation}
\Phi_{\text{dipole}}({\bf x}) = 
\Phi({\bf x}- {\bf e}\epsilon/2) - \Phi({\bf x}+{\bf e}\epsilon/2)
\label{Phidi}
\end{equation}
where $\epsilon$ is the width and ${\bf e}$ the unit vector in the
direction of the displacement. In the following we will assume that
both the displacement and the forces show in the $x$-direction. 
Using the asymptotic expression (\ref{Phias}) and taking the leading
contribution with respect to the distance $\epsilon$ one obtains:
\begin{equation}
\Phi^{\text{as}}_{\text{dipole}}= \epsilon \frac{y x}{8 \pi r^4}z(1-z)
\end{equation}
The velocity field then becomes asymptotically:
\begin{eqnarray}
v_x^{as}({\bf x}) & = &  \frac{\epsilon x}{8 \pi r^4} \left( 1-4
  \frac{y^2}{r^2} \right ) z (1-z) \nonumber \\
v_y^{as}({\bf x}) & = & -\: \frac{\epsilon y}{8 \pi r^4} \left( 1-4
  \frac{x^2}{r^2} \right ) z (1-z) \nonumber \\ 
v_z({\bf x}) & = & 0
\label{vdip}
\end{eqnarray}
Note that the near field has a less pronounced, logarithmic singularity. 
The full two-dimensional vector-field $v_x({\bf r},0.5),v_y({\bf r},0.5))$ 
that follows from the
eqs.(\ref{vPhi},\ref{Phis},\ref{cnf},\ref{Phidi}) is shown for a pair
of opposite line force pointing in the positive and negative
$x$-direction at ${\bf r} = (\pm 1,0)$ in  
Fig.\ref{fig:flow}. In this case the fluid is sucked along the $y$-axis 
and pushed out in the $\pm x$-direction generating an eddy in
each quadrant close to the dipole.  
  
\subsection{Motion of advected particles}
We consider the motion of a small particle that is advected by the flow field
(\ref{vdip}). Its trajectory agrees with the streamlines of the vector field
${\bf v}({\bf x})$ which are the solutions of
\begin{equation}
\label{dx}
\dot{{\bf x}}(t) = {\bf v}({\bf x}(t))
\end{equation}
Because the velocity component in the $z$ direction vanishes the
 motion of a particle is strictly confined to the initial plane $z=
 z(0)$  as long as diffusion can be neglected. 
 Fig.~\ref{fig:flow} shows the form that an initially circular spot 
in the central plane ($z = 1/2$)
 acquires at four later times. We note 
 that the area covered by the spot
 does not change because the vector field has vanishing
 divergence. The apparent change of the area of a dye spot which is
 observed in the experiment is due to the fact that different layers
 that move differently are superimposed. The alternating
 stripes of colored and uncolored fluid that evolve after
 sufficiently long time within each vertical layer cover each other
 and therefore cannot easily be seen in the
 experiment. In the course of time these layers become
 ever thinner and eventually diffusion becomes effective.      
Around the center of each of the four eddies regions exist that
 remain invariant under the flow, i.e.  there is no transport out of
 these regions or into them other than diffusion.  

Finally we consider 
a particle that initially sits in the middle position
between the plates on the $x$-axis at a
distance $x_0$ from the origin. It moves with the fluid strictly in the
$x$-direction
according to
\begin{eqnarray}
\frac{d}{dt}x(t) & = & v_x(x(t),0,0.5) \nonumber \\
& = & - \frac{c}{x(t)^3}
\label{xad}
\end{eqnarray}
where $c=\epsilon/32 \pi$ is a constant. The solution of this
differential equation reads:
\begin{equation}
x(t) = \left ( x_0^4 + c t \right )^{1/4}
\label{xt}
\end{equation}
Here we have neglected the influence of diffusion. This will be of
primary importance for the motion in the $z$ direction because there
is no systematic force in this direction and, moreover, the
$z$-coordinate of the particle position strongly influences its
lateral velocity. A
diffusional motion in the $z$-direction averages over $z$ and results 
in an effective value of the constant $c$ which becomes smaller by the
factor 2/3.

\par
\begin{figure}
\includegraphics[width=8.4cm]{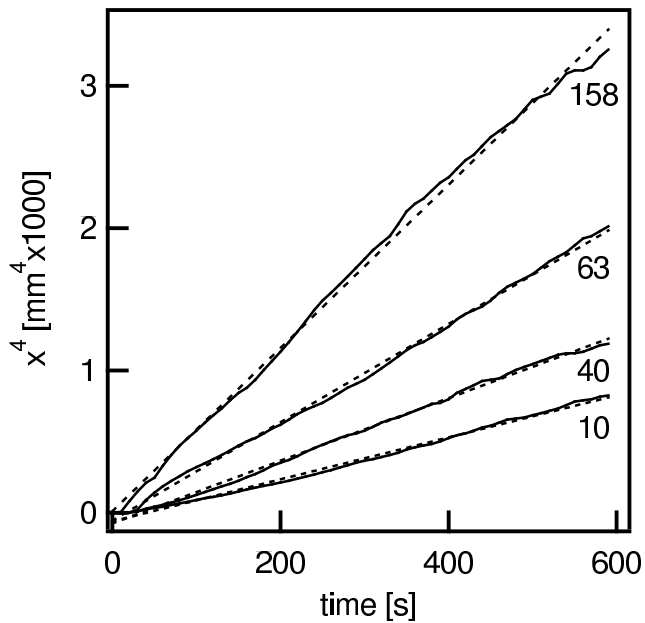}
\caption{The fourth power of the front positions (solid line) for four
  values of RF-power 
(10, 40, 63, 158 mW from bottom to top) of the dye spot
 indicated by the arrow in Fig.~\ref{fig:schmetterl2} measured at different instants
 of time (see also from Fig.~\ref{fig:distance}) falls on  straight
 lines as  predicted by the asymptotic law
  given in eq.~(\ref{xt}). Only
  for the very first points and at high RF-power deviations appear. From the
  inclination of the linear fits (broken lines) one obtains a reliable fit of the
  constant $c$ that determines the velocity field, see
  eq.~(\ref{xad}).
\label{fig:theory}}
\end{figure}
\begin{figure}
\includegraphics[width=8.4cm]{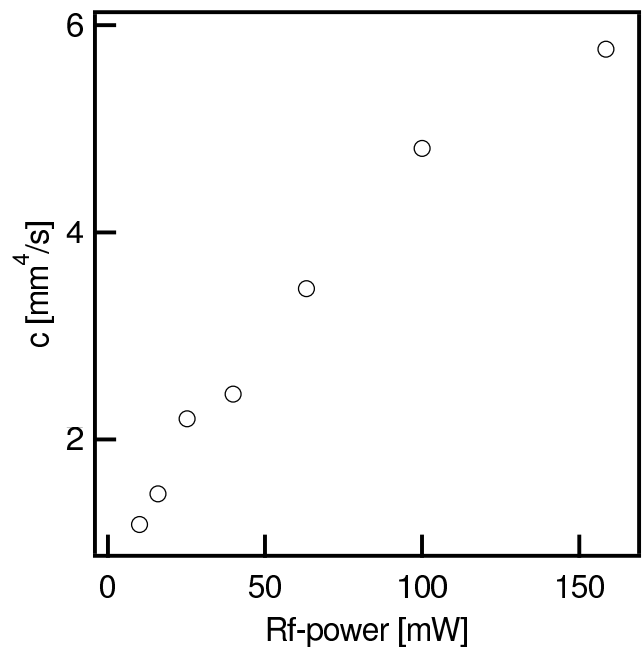}
\caption{The constant $c$ as determined from measurements of the
 position of the dye front is shown for different values of the
  RF-power.}
\label{fig:konstante}
\end{figure}

In order to compare with the theoretical prediction of eq.~(\ref{xt}) in Fig.~\ref{fig:theory}
the fourth power of the position is plotted against time for four
different RF-power values. 
Except for
the highest power value where the asymptotic velocity
field eq.~(\ref{xad}) might not be fully applicable, the experimental data
nicely fall onto straight lines. From the slope of this line we obtain
the constant $c$ which determines the velocity field, eq.~(\ref{xad}). The
value of $c$ increases with the applied RF-power, see Fig.~\ref{fig:konstante}. An
extrapolation of a 
linear fit of the data, however, would lead to a spurious finite positive
value of $c$ and consequently to finite velocities in the absence of driving. 
We therefore conclude that the increase of $c$ with the RF-power
must be weaker than linear.

\section{Discussion}

We applied fluorescence microscopy and FCS for the flow profile analysis of a SAW transducer 
in a 200$\mu$m thin fluid film. With the first technique the growth 
of the area of a Fluorescein spot was observed. From this the flow speed 
in x-direction could be determined. 

In the FCS experiments, the flow speed in a homogeneous dispersion of 
fluorescent beads was measured at different points in a 5mm wide channel. 
With this technique no limitations exist for the choice of the position
where the data is collected. 

Both methods yield a fast decay of the 
flow speed with distance from the transducer with a comparable rate. 
As a theoretical model, the Navier Stokes equations were solved for an
infinite fluid layer in the Stokes limit of low Reynolds numbers. The
conversion of the sound wave into a steady flow was described in terms
of body forces that were further simplified as line forces. Because
the actual spatial extension of the body forces is confined to a
narrow regime this seems to be a reasonable approximation. 
The resulting flow field is in good qualitative agreement with the
experimental findings. According to the theoretical prediction, the
velocity field asymptotically decays with the
third inverse power of the distance from the IDT. The resulting motion
of the front of a dye spot in the symmetry direction is in
quantitative agreement with the measurements.  

 
To determine the distance at which the influence of the transducer 
on the dye molecules vanishes, the growth speed ($v_A$) of the fluorescein 
covered area was measured. Plotting this data against the actual position 
of the advancing front shows also a fast decay until the diffusion 
constant of the the dye is reached. This distance 
depends logarithmically on the RF-power and is about 8 mm for the maximum 
applied 158 mW in the microscope study.

When the flow speed is measured at different points in the streamlines with both
methods, a linear behavior dependent on the RF-power can be observed.
However, when the constant $c$ in eq.~(\ref{xt}) is determined from the
$x$ position vs time data for different Hf-values, no line through the origin can be fitted.
So following eq.~(\ref{xad}) the flow speed cannot depend linearly on the power
at a certain point in the flow field. The power range where measurements where taken,   
therefore probably represents an approximately linear regime.

Finally we address the question what one can learn from the presented
experiments and theory in order to accelerate the hybridization of DNA in a thin fluid
film with the lateral size of a normal glass slide (75x25mm).  
For this size, diffusive transport becomes extremely slow and should be
complemented by advected transport. The flow induced by surface
acoustic waves can be efficiently used for this purpose. 
However, because of the
rapid decay of the induced velocity field away from the transducer the
maximal mixing range was found to have a linear dimension of about 8
mm in the present experiment at the highest RF-power. A further
increase of the power does not seem feasible because it would enlarge 
the mixing range only insignificantly.
To account for the limited range of the induced flow by the SAW transducer
the number of transducers on a slide should be 
increased. In the calculated flow field in Fig.~\ref{fig:flow} four eddies appear
close to the transducer where particles can be trapped. To avoid this
problem, pairs of transducers should always be operated alternately.
So the trapped particles from one transducer will be moved by the other
and vice versa.

\section*{Acknowledgment}
The authors gratefully acknowledge financial support by the Deutsche
Forschungsgemeinschaft (DFG) under the Sonderforschungsbereich 486 and
the Bayrische Forschungsstiftung under the Programme
``ForNano''.



\end{document}